\documentclass[journal=jacsat,manuscript=article]{achemso}

\usepackage[version=3]{mhchem} 
\usepackage{amsmath,amssymb,mathrsfs,accents}
\usepackage{color}
\usepackage[scr=rsfso]{mathalpha}
\usepackage{booktabs}                   
\usepackage{multirow}                   
\usepackage{diagbox}                    
\usepackage{bm}
\usepackage{physics}
\usepackage[T1]{fontenc}
\usepackage[utf8x]{inputenc}

\author{Yu-Chen Wei}
\affiliation{Institute of Atomic and Molecular Sciences, Academia Sinica, Taipei 106, Taiwan}
\alsoaffiliation{Department of Chemistry, National Taiwan University, Taipei 106, Taiwan}
\author{Liang-Yan Hsu}
\affiliation{Institute of Atomic and Molecular Sciences, Academia Sinica, Taipei 106, Taiwan}
\alsoaffiliation{Department of Chemistry, National Taiwan University, Taipei 106, Taiwan}
\alsoaffiliation{National Center for Theoretical Sciences, Taipei 106, Taiwan}
\email{lyhsu@gate.sinica.edu.tw}

\title[An \textsf{achemso} demo]
  {Wide-Dynamic-Range Control of Quantum-Electrodynamic Electron Transfer Reactions in the Weak Coupling Regime}

\begin{document}








\begin{abstract}
  Catalyzing reactions effectively by vacuum fluctuations of electromagnetic fields is a significant challenge within the realm of chemistry. Different from most studies based on vibrational strong coupling, we introduce an innovative catalytic mechanism driven by weakly coupled polaritonic fields. Through the amalgamation of macroscopic quantum electrodynamics (QED) principles with Marcus electron transfer (ET) theory, our results reveal that ET reaction rates can be precisely modulated across a wide dynamic range by controlling the size and structure of nanocavities. Comparing to QED-driven radiative ET rates in free space, plasmonic cavities induce substantial rate enhancements spanning from orders of magnitude ranging from $10^3$-fold to $10^1$-fold. By contrast, Fabry-Per\'{o}t cavities engender rate suppression spanning from $10^{-2}$-fold to $10^{-1}$-fold. This work overcomes the necessity of using strong light-matter interactions in QED chemistry, opening up a new era of manipulating QED-based chemical reactions in a wide dynamic range.
\end{abstract}

Quantum-electrodynamics (QED) chemistry has become an active field in the past few years due to its potential applications in chemical reactions and optoelectronics \cite{nagarajan2021chemistry,ebbesen2016hybrid,garcia2021manipulating,li2022molecular,george2023polaritonic}. Recent studies have shown that the formation of hybrid light-matter states in the strong coupling regime can modify chemical reactions \cite{nagarajan2021chemistry,ebbesen2016hybrid}. Experimentally, the hybrid light-matter states can be utilized to modify the molecular self-assembly \cite{2021supramolecular,hirai2021selective,sandeep2022manipulating}, alter the reaction rates \cite{thomas2016ground,vergauwe2019modification,ahn2023modification} and selectively control the reaction pathways \cite{thomas2019tilting,pang2020role,sau2021modifying}. In addition, several theoretical and experimental studies have proposed that it is possible to trigger collective chemical reactions by strongly coupling the molecular ensemble to a confined photon mode \cite{herrera2016cavity,lather2019cavity,sidler2020polaritonic,li2021collective,gomez2023vibrational,du2023vibropolaritonic}. Among the various demonstrations, chemical reactions triggered by 
formations of hybrid light-matter states require strong light-matter coupling between electromagnetic (EM) modes and molecular transitions. However, maintaining the condition of strong coupling proves challenging due to the constant shifts in the environmental dielectric function during chemical reactions \cite{ebbesen2016hybrid,nagarajan2021chemistry,simpkins2021mode}. In addition, the reported dynamic ranges of tuning reaction rates via strong vibrational coupling have been found to be limited to less than one order of magnitude \cite{nagarajan2021chemistry,garcia2021manipulating,ahn2023modification}. These issues underscore the limitations in controlling chemical reactions via strong light-matter coupling. 

Different from the previous approaches, Semenov et~al. proposed that a single cavity mode remarkably enhances electron transfer (ET) rates in the inverted Marcus regime   \cite{semenov2019electron}. In addition, Wei et~al. demonstrated that ET reaction rates can be modulated by infinite photonic modes, even in the absence of strong light-matter coupling  \cite{wei2022cavity}. These insightful studies inspire us to ask a question: Is it possible to effectively control chemical reaction rates over several orders of magnitude without relying on strong light-matter coupling? To answer this question, we present a feasible weak-coupling approach to manipulate ET rates via tuning cavity sizes. To quantitatively describe QED effects on ET reactions in a medium, we develop a macroscopic QED version of ET (mQED-ET) theory by considering the influence of infinite polaritonic modes. According to this approach, we study how vacuum fluctuations of infinite polaritonic modes influence the radiative charge recombination process in a donor-acceptor dyad. This study offers new insights into the underlying mechanisms and experimental design principles for QED-controlled chemical reactions in the weak light-matter coupling regime.

\section{Theoretical Framework}
To properly describe the light-matter interaction in a dispersive and absorbing dielectric environment, we start from the length-gauge Hamiltonian in the framework of macroscopic QED \cite{gruner1996green,dung1998three,buhmann2013dispersion}. 

\begin{align}
\label{Eq:mininal coupling_H_GM}
\hat{H}_\mathrm{GM}=&~\sum_\alpha\frac{\hat{\mathbf{p}}_\alpha^2}{2m_\alpha}+\frac{\mathrm{1}}{2}\int d\mathbf{r}\hat{\rho}_{\mathrm{M}}(\mathbf{r})\cdot\hat{\phi}_{\mathrm{M}}(\mathbf{r})\nonumber\\
&\hspace{0.3cm}+\int d\mathbf{r}\int_0^\infty d\omega\hbar\omega \hat{\mathbf{f}}^\dag(\mathbf{r},\omega)\cdot\hat{\mathbf{f}}(\mathbf{r},\omega)-\hat{\boldsymbol{\mu}}\cdot\hat{\mathbf{E}}(\mathbf{r}_\mathrm{M})+\frac{\vert\hat{\boldsymbol{\mu}}\vert^2}{2\epsilon_0 V_\mathrm{eff}}
\end{align}
$m_\alpha$ and $\hat{\mathbf{p}}_\alpha$ indicate the mass and the momentum operator of the $\alpha^\mathrm{th}$ particle, respectively. The symbol $\hat{\rho}_{\mathrm{M}}(\mathbf{r})$ represents a charge density operator for the molecule. The symbols $\hat{\phi}_{\mathrm{M}}(\mathbf{r})$ and $\hat{\phi}(\mathbf{r})$ correspond to molecular and external Coulomb potential operators, respectively. $\hat{\mathbf{f}}^\dag(\mathbf{r},\omega)$ ($\hat{\mathbf{f}}(\mathbf{r},\omega)$) is the creation (annihilation) operator for bosonic vector fields (polariton) in macroscopic QED.\cite{gruner1996green,dung1998three,buhmann2013dispersion} The term $\hat{\boldsymbol{\mu}}\cdot\hat{\mathbf{E}}(\mathbf{r}_\mathrm{M})$ describes the light-matter interaction in the electric-dipole form, where $\hat{\boldsymbol{\mu}}$ and $\hat{\mathbf{E}}(\mathbf{r}_\mathrm{M})$ represent the dipole moment operator and the electric field operator, respectively. The last term is the
dipole self-energy, where $V_\mathrm{eff}$ is the effective mode volume of the transverse EM fields. The detailed discussion about Eq.~\ref{Eq:mininal coupling_H_GM} is shown in our previous study \cite{wei2023polaritonic}. Note that the length gauge Hamiltonian retains intramolecular Coulomb interactions and electromagnetic vacuum fluctuations caused by the transversal polaritonic degrees of freedom in medium. The former allows us to describe the Coulomb electronic coupling and the vibronic coupling in Marcus theory, and the latter enables us to depict the medium-assisted QED effects on ET processes.

To obtain the effective electronic coupling, we develop the model Hamiltonian considering electronic, vibraltional and polaritonic degrees of freedom [see Supplementary Information Section 1], adopt the unitary transformation along polaritonic coordinates [see Supplementary Information Section 2] and the small-polaron transformation [see Supplementary Information Section 3]. Next, to derive a Marcus-type expression for the ET rates, we apply the Fermi’s golden rule and evaluate the ET rates in the electronic weak-coupling regime based on the same approximations in QED-ET theory, including the Condon approximation, the short-time approximation, and the low-vibrational-frequency limit \cite{wei2022cavity}. The details of how to derive the mQED-ET theory from the Fermi’s golden rule can be found in Supplementary Information Section 4. Notably, due to the slight effects of polaritonic displacements on light-matter coupling strength \cite{wei2023polaritonic}, the effects of light–matter coupling induced by permanent dipoles and dipole self-energy are neglected here. Through the above processes, we derive an explicit expression for the total ET rate $k_\mathrm{ET}$ as
\begin{align}
\label{Eq:def_kET}
k_\mathrm{ET}=&~k_\mathrm{Marcus}+k_\mathrm{QED},
\end{align}
where $k_\mathrm{Marcus}$ corresponds to  the Marcus nonradiative ET rate \cite{marcus1993electron}. $k_\mathrm{QED}$ is the QED-driven radiative ET rate, which can be expressed as the integral of polaritonic spectral density $J_\mathrm{pol}(\omega)$ and vibrational transition density of states (DOS) $\rho_\mathrm{vib}(\omega)$ in the following form:
\begin{align}
\label{Eq:kET_QED}
    k_\mathrm{QED}=~&\frac{2\pi}{\hbar}\int_0^\infty\mathrm{d}\omega~J_\mathrm{pol} (\omega)\times\rho_\mathrm{vib}(\omega),
\end{align}
where $J_\mathrm{pol}(\omega)$ is associated with molecular ET transition dipole and local polaritonic DOS (see exact expression of $J_\mathrm{pol}(\omega)$ in Supplementary Information Section 4.2).

\begin{figure}[hbtp!]
\centering
    \includegraphics[width=\textwidth]{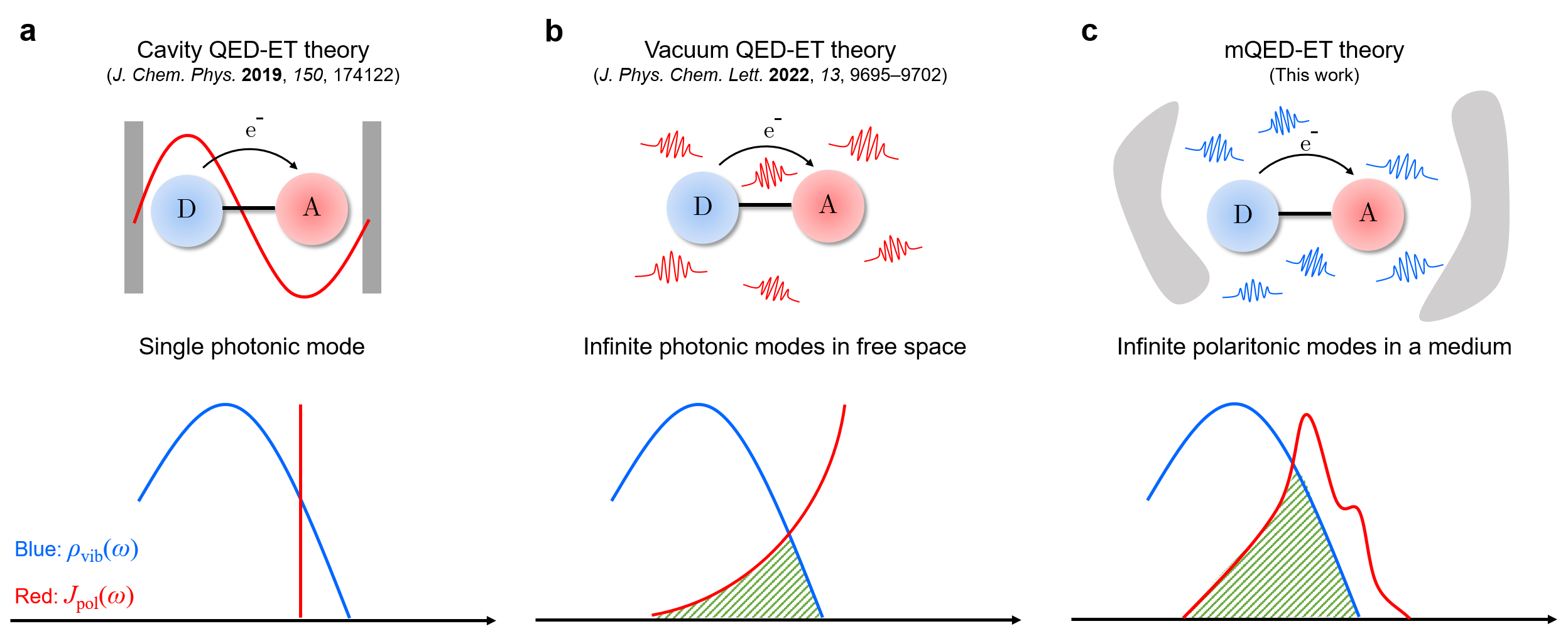}
    \caption{Schematic illustration of  spectral overlaps for the radiative ET process. $\mathbf{a}$  Cavity QED-ET theory developed by Semenov and Nitzan \cite{semenov2019electron}. $\mathbf{b}$ Vacuum QED-ET theory developed by Wei and Hsu \cite{wei2022cavity}. $\mathbf{c}$ mQED-ET theory in this work. The blue curve represents the vibrational transition density of states $\rho_\mathrm{vib}(\omega)$. The red curve represents the photonic/polaritonic spectral density $J_\mathrm{pol}(\omega)$. The green diagonal stripes represent the concept of electron-transfer overlap, i.e., the spectral overlap between vibrational transition density of states and the polaritonic spectral density. The x axis indicates the photonic frequency.}
    \label{Fig:1}
\end{figure}

 Note that the integral of $J_\mathrm{pol}(\omega)$ corresponds to the square of the light–matter coupling strength and $\rho_\mathrm{vib}(\omega)$ describes vibronic transitions during an ET process \cite{wei2022cavity}. According to Eq.~\ref{Eq:kET_QED}, we propose a generalized version of ``electron-transfer overlap'' corresponding to the spectral overlap of the polaritonic spectral density $J_\mathrm{pol}(\omega)$ (the red curve) and the vibrational transition DOS $\rho_\mathrm{vib}(\omega)$ (the blue curve), whose physical pictures are shown in Fig.~\ref{Fig:1}. If we consider the condition of a single photonic mode, the photonic DOS behaves as a single-peak delta function (Fig.~\ref{Fig:1}a) and Eq~\ref{Eq:def_kET} can be reduced to the cavity QED version of ET rates, as derived by Semenov and Nitzan under the condition of slow electron and fast cavity mode \cite{semenov2019electron} (see Supplementary Information Section 5). Next, in the case of a dielectric environment corresponding to free space, the infinite photonic modes exhibit a cubic frequency dependence in the photonic DOS (Fig.~\ref{Fig:1}b), aligning with the vacuum QED-ET theory (see Supplementary Information Section 6) \cite{wei2022cavity}. In this work, $J_\mathrm{pol}(\omega)$ can describe the polaritonic DOS in an arbitrary dielectric environment (Fig.~\ref{Fig:1}c), encompassing the cavity QED-ET theory and the vacuum QED-ET theory as its special cases. 

The niches of Eq.~\ref{Eq:kET_QED} are elaborated as follows. First, $J_\mathrm{pol}(\omega)$ can be evaluated in an arbitrary dielectric environment without free parameters based on analytical methods (e.g. the Fresnel method and the Mie theory) or computational electrodynamics packages \cite{wang2020coherent,lee2020controllable,wei2021can}. This feasibility can help experimentalists to predict and analyze $k_\mathrm{QED}$. Second, Eq.~\ref{Eq:kET_QED} suggests that the manipulation of ET rates in the weak light-matter coupling regime is possible by adjusting electron-transfer overlap, which can be achieved by modifying cavity structures during experiments. In simpler terms, the mQED-ET theory can be used by experimentalists to design photonic/plasmonic structures for controlling QED-driven chemical reactions.

\section{Weak-Coupling Control of Molecular ET Rates}

\begin{figure}[hbtp!]
    \includegraphics[width=\textwidth]{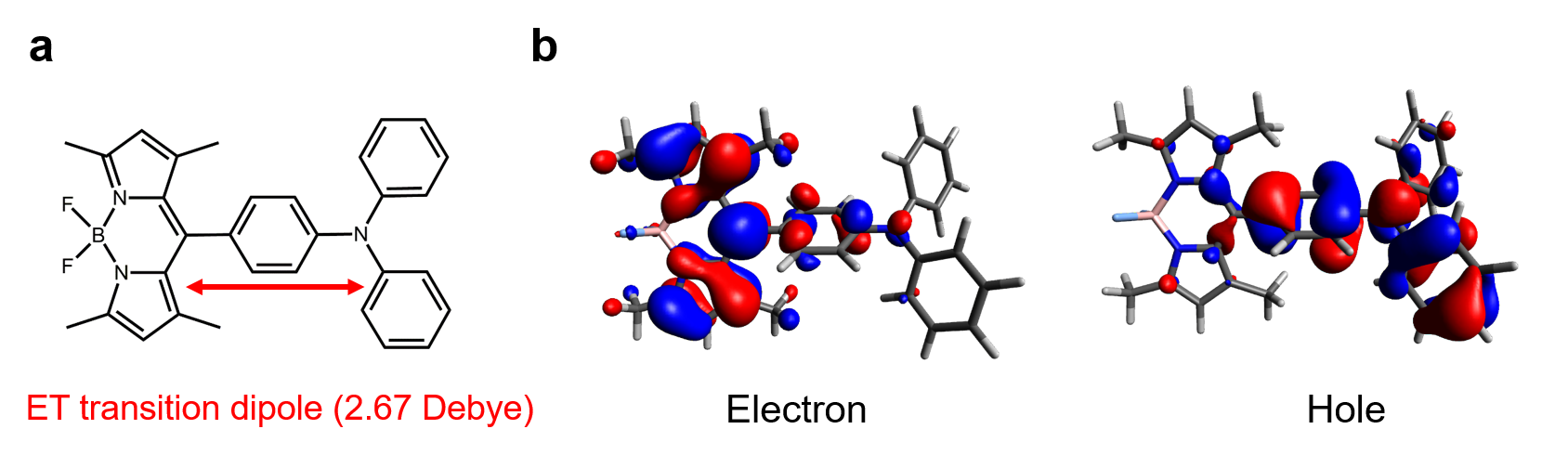}
    \caption{Candidate molecule for controlling ET reaction via weak light-matter interaction. $\mathbf{a}$ Chemical structure of BODIPY-TPA. $\mathbf{b}$ Natural transition orbital analysis of the donor-acceptor dyad BODIPY-TPA. The red double arrow in ($\mathbf{a}$) represents the direction of ET transition dipole of the charge-transfer transition. The molecular geometry in ($\mathbf{b}$) is optimized in the charge-transfer state. See Methods for the detailed information of the quantum chemical calculation.}
    \label{Fig:2}
\end{figure}

In order to manipulate molecular ET rates via QED effects, it is essential for the QED-driven ET rates $k_\mathrm{QED}$ to be significantly higher than the Marcus nonradiative ET rates $k_\mathrm{Marcus}$. According to the QED-ET theory, such situation usually arises in the highly inverted regime \cite{wei2022cavity}. Guided by this principle, our investigation focuses on examining the impact of QED effects on the process of radiative charge recombination in a donor-acceptor dyad BODIPY-TPA based on the fluorescent BODIPY functionalized with the triphenylamine (TPA) (Fig.~\ref{Fig:2}a). In terms of the charge recombination,  the electron density undergoes migration from the BODIPY moiety to the TPA moiety, as revealed by the natural transition orbital analysis (Fig.~\ref{Fig:2}b). It is noteworthy that the remarkable photoluminescent quantum yield (PLQY = 12.5\%) supports the occurrence of a substantial radiative ET process in the BODIPY-TPA system \cite{buck2019intramolecular}. This aspect makes it particularly well-suited for manipulation through light-matter interactions.

\begin{figure}[hbtp!]
    \includegraphics[width=\textwidth]{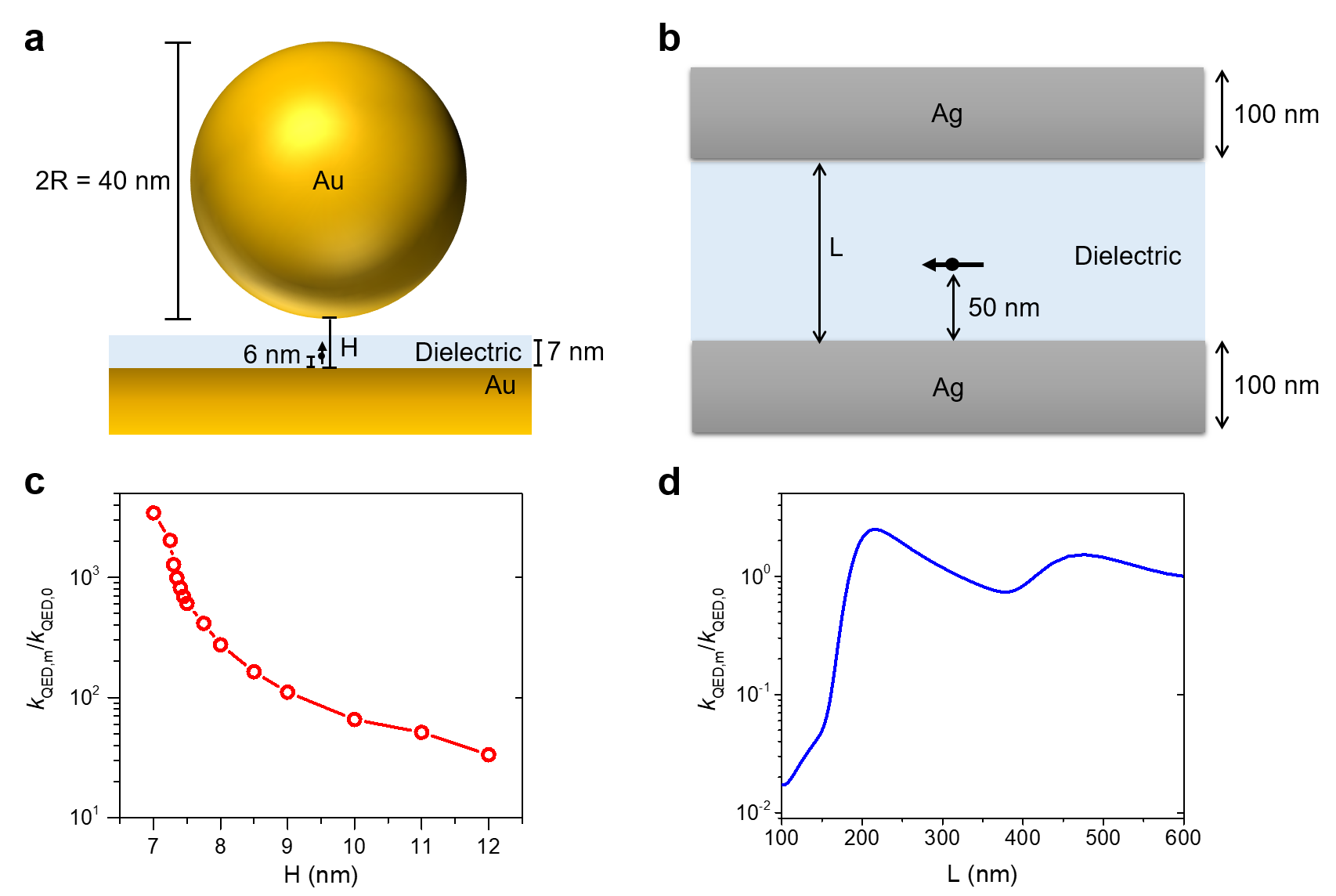}
    \caption{Manipulating QED-driven ET rates via modifying cavity structures. $\mathbf{a}$ Schematic illustration of  the plasmonic cavity. $\mathbf{b}$ Schematic illustration of the photonic cavity. The black arrow indicates the position and the direction of the molecular transition dipole. H represents the gap between the metal surface and the nanosphere. L indicates the cavity length. The dielectric layers in ($\mathbf{a}$) and ($\mathbf{b}$) are non-absorbing with a constant refractive index $n = 1.5$. $\mathbf{c}$ QED-driven ET rate enhancements in the plasmonic cavity. $\mathbf{d}$ QED-driven ET rate enhancements in the photonic cavity. $k_\mathrm{QED,m}$ and $k_\mathrm{QED,0}$ indicate the radiative ET rates in a medium and free space, respectively. Note that the hollow circles in ($\mathbf{c}$) represents the data points calculated by the computational electrodynamics packages, while the blue line in $k_\mathrm{QED,0}$ are evaluated by the analytical method  \cite{wei2022cavity}. See Methods for the detailed information of evaluating $k_\mathrm{QED}$.}
    \label{Fig:3}
\end{figure}

Furthermore, to regulate charge recombination rates under the weak light-matter coupling regime, two types of cavities are designed. The first type is a plasmonic cavity comprising a dielectric-coated gold surface and a gold nanosphere (Fig.~\ref{Fig:3}a). The second type is a photonic cavity, specifically a Fabry-Perot cavity, constructed by sandwiching a dielectric layer between two thin films of silver (Fig.~\ref{Fig:3}b). Note that the two kinds of cavities mentioned above are are widely used in the fields of nanophotonics, plasmonics and polariton chemistry for developing significant light-matter interaction \cite{vasa2018strong,garcia2021manipulating,rivera2020light}. In addition, molecules in these cavities are under weak light-matter coupling regime owing to the two reasons. First, the molecular transition energy greatly deviates from the peak energy of the $J_\mathrm{pol}(\omega)$ (Fig. S1), leading to the poor ET spectral overlap. Second, the photonic dissipation estimated from the peak width of $J_\mathrm{pol}(\omega)$ majorly exceeds the light-matter coupling strength (Fig. S1). The detailed method of evaluating light-matter coupling strength is shown in our previous study \cite{wei2023polaritonic}. Worth to mention, the energy gap dependence of the ET rates in the two nanocavities reveals that $k_\mathrm{QED}$ significantly surpasses $k_\mathrm{Marcus}$ in the charge recombination process of BODIPY-TPA (Fig. S2), supporting the rationale for disregarding $k_\mathrm{Marcus}$. To modify QED-driven ET rates, adjustments are made to the gap distance (H) between the metal surface and the nanosphere in the plasmonic cavity, as well as the length (L) of the cavity in the photonic cavity.  Based on the mQED-ET theory, the rate enhancements $k_\mathrm{QED,m}/k_\mathrm{QED,0}$ are evaluated, where $k_\mathrm{QED,m}$ and $k_\mathrm{QED,0}$ represent the radiative ET rates in a medium and vacuum, respectively (Fig.~\ref{Fig:3}c and~\ref{Fig:3}d). The results demonstrate that the plasmonic cavity causes ET rate enhancements ranging from $10^3-10^1$, whereas the photonic cavity suppresses the ET rates by a factor of $10^{-2}-10^{-1}$. The dynamic ranges of modulating radiative ET rates in both nanocavities reach an impressive two orders of magnitude, which far exceeds the rate change induced by vibrational strong coupling \cite{nagarajan2021chemistry,garcia2021manipulating,ahn2023modification}. Such wide tunability of QED-driven ET rates holds significant potential for their practical application in QED-based chemical reactions.

\begin{figure}[hbtp!]
\centering
    \includegraphics[width=0.6\textwidth]{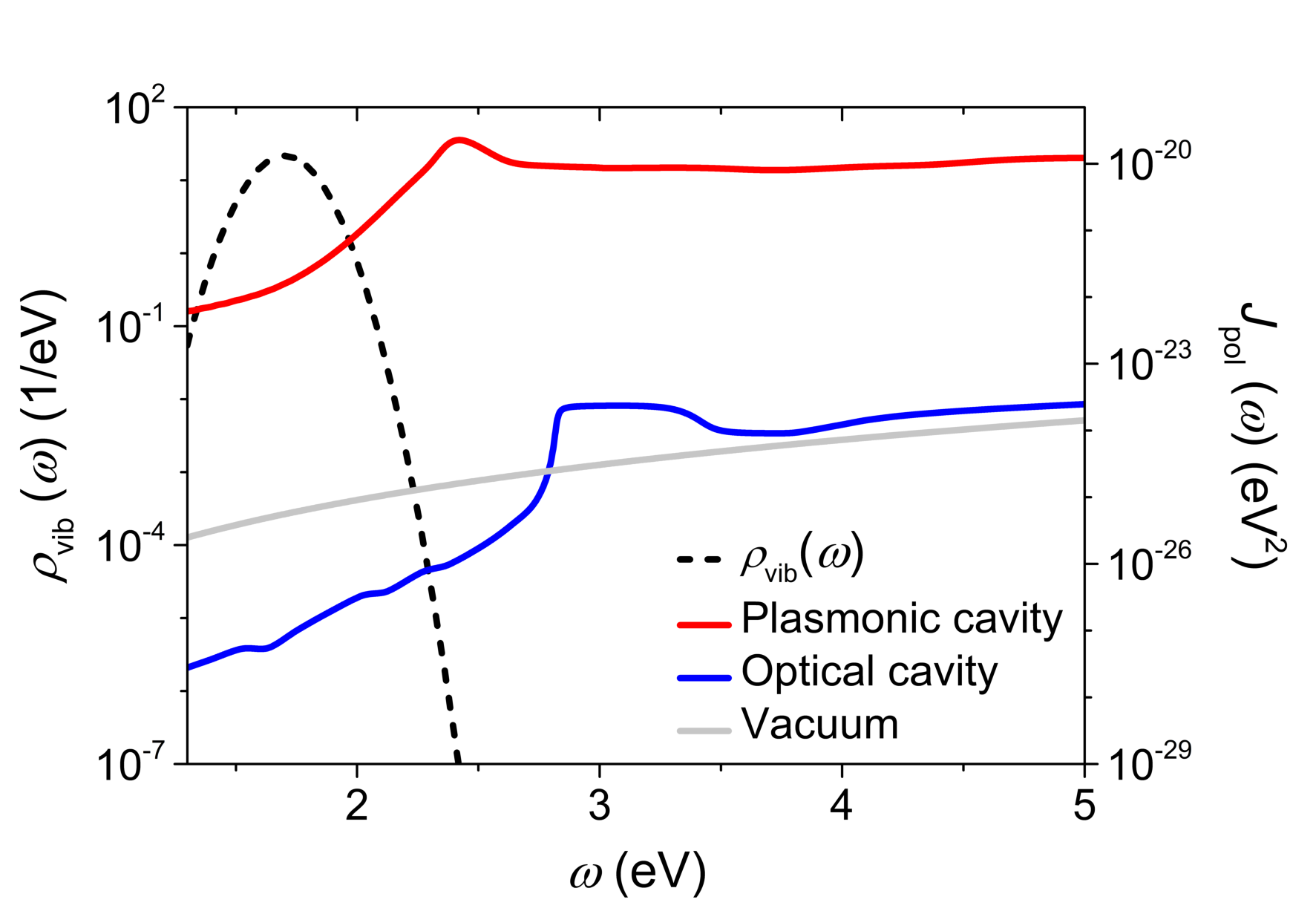}
    \caption{Electron-transfer overlap in the plasmonic and photonic cavities. The following parameters for the two cavities are applied: $\mathrm{H}=7$ nm and $\mathrm{L}=100$ nm. The parameters of $\rho_\mathrm{vib}(\omega)$ follow the ones in Figure~\ref{Fig:3}.}
    \label{Fig:4}
\end{figure}

The above phenomena can be explained through the analysis of the electron-transfer overlap (Fig.~\ref{Fig:4}). In the plasmonic cavity, the significant increase in QED-driven ET rates is attributed to the large polaritonic spectral density $J_\mathrm{pol}(\omega)$ contributed by surface plasmon polaritons. On the other hand, the photonic cavity mainly allows the existence of polaritonic modes with wavelengths associated with the cavity length. This leads to a reduction in the $J_\mathrm{pol}(\omega)$ and consequently a decrease in the corresponding QED-driven ET rates. Note that the results align with the Purcell effects, as both radiative electron transfer (ET) rates and spontaneous emission rates are influenced by the local polaritonic/photonic DOS \cite{purcell1946proceedings,vahala2003optical}. Furthermore, akin to spontaneous emission, the polariton-assisted ET rates also exhibit a near absence of activation energy (Fig. S3), attributed to a mechanism analogous to that explained in the QED-ET theory  \cite{wei2022cavity}.

Our key findings are summarized as follows. First, the manipulation of chemical reaction rates through weak light-matter interaction has been demonstrated. This approach is believed to be more practical for experimentalists in terms of leveraging QED effects to control chemical reactions compared to previous methods that require strong light-matter coupling conditions. Second, we have derived a mQED-version ET theory and generalized the concept of electron-transfer overlap: the spectral overlap of the polaritonic spectral density $J_\mathrm{pol}(\omega)$ and vibrational transition DOS $\rho_\mathrm{vib}(\omega)$, which elaborates the interplay of polaritonic modes and ET reactions. This breakthrough allows us to quantify QED-driven ET rates in an arbitrary absorbing and dispersive medium. Third, our theory predicted that different types of cavities cause opposite effects on molecular charge recombination rates. The underlying mechanism can be understood through the physical picture of electron-transfer overlap. Fourth, the QED-driven ET rates show great adjustability via tuning sizes of nanocavities, giving rise to the remarkable dynamic ranges spanning two orders of magnitude within the designed cavities. Note that the wide dynamic range shown in weak-coupling QED-ET reactions surpasses the range observed in chemical reactions controlled by vibrational strong coupling \cite{nagarajan2021chemistry,garcia2021manipulating,ahn2023modification}. This exceptional capability to manipulate ET rates possesses substantial potential for controlling chemical reactions through QED effects. In summary, this study offers a practical approach to effectively control chemical reactions via weak light-matter interactions. We anticipate that this work will inspire and support novel developments of QED-based chemical reactions.

\begin{acknowledgement}

We thank Academia Sinica (AS-CDA-111-M02) and the Ministry
of Science and Technology of Taiwan (MOST 111-2113-M-001-027-MY4) for generous support.

\end{acknowledgement}

\begin{suppinfo}

The Supporting Information is available free of charge.

 Model Hamiltonian; Unitary transformation along polaritonic coordinates; Small-polaron transformation; Derivation of Equations 2 and 3; Conversion of mQED-ET Theory to Cavity QED-E Theory; Conversion of mQED-ET Theory to o Vacuum QED-ET Theory; Supplementary Figures.

\end{suppinfo}

\bibliography{achemso-demo}

\providecommand{\latin}[1]{#1}
\makeatletter
\providecommand{\doi}
  {\begingroup\let\do\@makeother\dospecials
  \catcode`\{=1 \catcode`\}=2 \doi@aux}
\providecommand{\doi@aux}[1]{\endgroup\texttt{#1}}
\makeatother
\providecommand*\mcitethebibliography{\thebibliography}
\csname @ifundefined\endcsname{endmcitethebibliography}
  {\let\endmcitethebibliography\endthebibliography}{}
\begin{mcitethebibliography}{37}
\providecommand*\natexlab[1]{#1}
\providecommand*\mciteSetBstSublistMode[1]{}
\providecommand*\mciteSetBstMaxWidthForm[2]{}
\providecommand*\mciteBstWouldAddEndPuncttrue
  {\def\EndOfBibitem{\unskip.}}
\providecommand*\mciteBstWouldAddEndPunctfalse
  {\let\EndOfBibitem\relax}
\providecommand*\mciteSetBstMidEndSepPunct[3]{}
\providecommand*\mciteSetBstSublistLabelBeginEnd[3]{}
\providecommand*\EndOfBibitem{}
\mciteSetBstSublistMode{f}
\mciteSetBstMaxWidthForm{subitem}{(\alph{mcitesubitemcount})}
\mciteSetBstSublistLabelBeginEnd
  {\mcitemaxwidthsubitemform\space}
  {\relax}
  {\relax}

\bibitem[Nagarajan \latin{et~al.}(2021)Nagarajan, Thomas, and
  Ebbesen]{nagarajan2021chemistry}
Nagarajan,~K.; Thomas,~A.; Ebbesen,~T.~W. Chemistry under vibrational strong
  coupling. \emph{J. Am. Chem. Soc.} \textbf{2021}, \emph{143},
  16877--16889\relax
\mciteBstWouldAddEndPuncttrue
\mciteSetBstMidEndSepPunct{\mcitedefaultmidpunct}
{\mcitedefaultendpunct}{\mcitedefaultseppunct}\relax
\EndOfBibitem
\bibitem[Ebbesen(2016)]{ebbesen2016hybrid}
Ebbesen,~T.~W. Hybrid light--matter states in a molecular and material science
  perspective. \emph{Acc. Chem. Res.} \textbf{2016}, \emph{49},
  2403--2412\relax
\mciteBstWouldAddEndPuncttrue
\mciteSetBstMidEndSepPunct{\mcitedefaultmidpunct}
{\mcitedefaultendpunct}{\mcitedefaultseppunct}\relax
\EndOfBibitem
\bibitem[Garcia-Vidal \latin{et~al.}(2021)Garcia-Vidal, Ciuti, and
  Ebbesen]{garcia2021manipulating}
Garcia-Vidal,~F.~J.; Ciuti,~C.; Ebbesen,~T.~W. Manipulating matter by strong
  coupling to vacuum fields. \emph{Science} \textbf{2021}, \emph{373},
  0336\relax
\mciteBstWouldAddEndPuncttrue
\mciteSetBstMidEndSepPunct{\mcitedefaultmidpunct}
{\mcitedefaultendpunct}{\mcitedefaultseppunct}\relax
\EndOfBibitem
\bibitem[Li \latin{et~al.}(2022)Li, Cui, Subotnik, and Nitzan]{li2022molecular}
Li,~T.~E.; Cui,~B.; Subotnik,~J.~E.; Nitzan,~A. Molecular polaritonics:
  chemical dynamics under strong light--matter coupling. \emph{Annu. Rev. Phys.
  Chem.} \textbf{2022}, \emph{73}, 43--71\relax
\mciteBstWouldAddEndPuncttrue
\mciteSetBstMidEndSepPunct{\mcitedefaultmidpunct}
{\mcitedefaultendpunct}{\mcitedefaultseppunct}\relax
\EndOfBibitem
\bibitem[George and Singh(2023)George, and Singh]{george2023polaritonic}
George,~J.; Singh,~J. Polaritonic Chemistry: Band-Selective Control of Chemical
  Reactions by Vibrational Strong Coupling. \emph{ACS Catalysis} \textbf{2023},
  \emph{13}, 2631--2636\relax
\mciteBstWouldAddEndPuncttrue
\mciteSetBstMidEndSepPunct{\mcitedefaultmidpunct}
{\mcitedefaultendpunct}{\mcitedefaultseppunct}\relax
\EndOfBibitem
\bibitem[Joseph \latin{et~al.}(2021)Joseph, Kushida, Smarsly, Ihiawakrim,
  Thomas, Paravicini-Bagliani, Nagarajan, Vergauwe, Devaux, Ersen,
  \latin{et~al.} others]{2021supramolecular}
Joseph,~K.; Kushida,~S.; Smarsly,~E.; Ihiawakrim,~D.; Thomas,~A.;
  Paravicini-Bagliani,~G.~L.; Nagarajan,~K.; Vergauwe,~R.; Devaux,~E.;
  Ersen,~O., \latin{et~al.}  Supramolecular assembly of conjugated polymers
  under vibrational strong coupling. \emph{Angew. Chem., Int. Ed.}
  \textbf{2021}, \emph{60}, 19665--19670\relax
\mciteBstWouldAddEndPuncttrue
\mciteSetBstMidEndSepPunct{\mcitedefaultmidpunct}
{\mcitedefaultendpunct}{\mcitedefaultseppunct}\relax
\EndOfBibitem
\bibitem[Hirai \latin{et~al.}(2021)Hirai, Ishikawa, Chervy, Hutchison, and
  Uji-i]{hirai2021selective}
Hirai,~K.; Ishikawa,~H.; Chervy,~T.; Hutchison,~J.~A.; Uji-i,~H. Selective
  crystallization via vibrational strong coupling. \emph{Chem. Sci.}
  \textbf{2021}, \emph{12}, 11986--11994\relax
\mciteBstWouldAddEndPuncttrue
\mciteSetBstMidEndSepPunct{\mcitedefaultmidpunct}
{\mcitedefaultendpunct}{\mcitedefaultseppunct}\relax
\EndOfBibitem
\bibitem[Sandeep \latin{et~al.}(2022)Sandeep, Joseph, Gautier, Nagarajan,
  Sujith, Thomas, and Ebbesen]{sandeep2022manipulating}
Sandeep,~K.; Joseph,~K.; Gautier,~J.; Nagarajan,~K.; Sujith,~M.; Thomas,~K.~G.;
  Ebbesen,~T.~W. Manipulating the Self-Assembly of Phenyleneethynylenes under
  Vibrational Strong Coupling. \emph{J. Phys. Chem. Lett.} \textbf{2022},
  \emph{13}, 1209--1214\relax
\mciteBstWouldAddEndPuncttrue
\mciteSetBstMidEndSepPunct{\mcitedefaultmidpunct}
{\mcitedefaultendpunct}{\mcitedefaultseppunct}\relax
\EndOfBibitem
\bibitem[Thomas \latin{et~al.}(2016)Thomas, George, Shalabney, Dryzhakov,
  Varma, Moran, Chervy, Zhong, Devaux, Genet, \latin{et~al.}
  others]{thomas2016ground}
Thomas,~A.; George,~J.; Shalabney,~A.; Dryzhakov,~M.; Varma,~S.~J.; Moran,~J.;
  Chervy,~T.; Zhong,~X.; Devaux,~E.; Genet,~C., \latin{et~al.}  Ground-state
  chemical reactivity under vibrational coupling to the vacuum electromagnetic
  field. \emph{Angew. Chem., Int. Ed.} \textbf{2016}, \emph{55},
  11462--11466\relax
\mciteBstWouldAddEndPuncttrue
\mciteSetBstMidEndSepPunct{\mcitedefaultmidpunct}
{\mcitedefaultendpunct}{\mcitedefaultseppunct}\relax
\EndOfBibitem
\bibitem[Vergauwe \latin{et~al.}(2019)Vergauwe, Thomas, Nagarajan, Shalabney,
  George, Chervy, Seidel, Devaux, Torbeev, and
  Ebbesen]{vergauwe2019modification}
Vergauwe,~R.~M.; Thomas,~A.; Nagarajan,~K.; Shalabney,~A.; George,~J.;
  Chervy,~T.; Seidel,~M.; Devaux,~E.; Torbeev,~V.; Ebbesen,~T.~W. Modification
  of enzyme activity by vibrational strong coupling of water. \emph{Angew.
  Chem., Int. Ed.} \textbf{2019}, \emph{58}, 15324--15328\relax
\mciteBstWouldAddEndPuncttrue
\mciteSetBstMidEndSepPunct{\mcitedefaultmidpunct}
{\mcitedefaultendpunct}{\mcitedefaultseppunct}\relax
\EndOfBibitem
\bibitem[Ahn \latin{et~al.}()Ahn, Triana, Recabal, Herrera, and
  Simpkins]{ahn2023modification}
Ahn,~W.; Triana,~J.; Recabal,~F.; Herrera,~F.; Simpkins,~B. Modification of
  ground state chemical reactivity via light-matter coherence in infrared
  cavities. \emph{Science} \emph{380}, 1165--1168\relax
\mciteBstWouldAddEndPuncttrue
\mciteSetBstMidEndSepPunct{\mcitedefaultmidpunct}
{\mcitedefaultendpunct}{\mcitedefaultseppunct}\relax
\EndOfBibitem
\bibitem[Thomas \latin{et~al.}(2019)Thomas, Lethuillier-Karl, Nagarajan,
  Vergauwe, George, Chervy, Shalabney, Devaux, Genet, Moran, \latin{et~al.}
  others]{thomas2019tilting}
Thomas,~A.; Lethuillier-Karl,~L.; Nagarajan,~K.; Vergauwe,~R.~M.; George,~J.;
  Chervy,~T.; Shalabney,~A.; Devaux,~E.; Genet,~C.; Moran,~J., \latin{et~al.}
  Tilting a ground-state reactivity landscape by vibrational strong coupling.
  \emph{Science} \textbf{2019}, \emph{363}, 615--619\relax
\mciteBstWouldAddEndPuncttrue
\mciteSetBstMidEndSepPunct{\mcitedefaultmidpunct}
{\mcitedefaultendpunct}{\mcitedefaultseppunct}\relax
\EndOfBibitem
\bibitem[Pang \latin{et~al.}(2020)Pang, Thomas, Nagarajan, Vergauwe, Joseph,
  Patrahau, Wang, Genet, and Ebbesen]{pang2020role}
Pang,~Y.; Thomas,~A.; Nagarajan,~K.; Vergauwe,~R.~M.; Joseph,~K.; Patrahau,~B.;
  Wang,~K.; Genet,~C.; Ebbesen,~T.~W. On the role of symmetry in vibrational
  strong coupling: the case of charge-transfer complexation. \emph{Angew.
  Chem., Int. Ed.} \textbf{2020}, \emph{59}, 10436--10440\relax
\mciteBstWouldAddEndPuncttrue
\mciteSetBstMidEndSepPunct{\mcitedefaultmidpunct}
{\mcitedefaultendpunct}{\mcitedefaultseppunct}\relax
\EndOfBibitem
\bibitem[Sau \latin{et~al.}(2021)Sau, Nagarajan, Patrahau, Lethuillier-Karl,
  Vergauwe, Thomas, Moran, Genet, and Ebbesen]{sau2021modifying}
Sau,~A.; Nagarajan,~K.; Patrahau,~B.; Lethuillier-Karl,~L.; Vergauwe,~R.~M.;
  Thomas,~A.; Moran,~J.; Genet,~C.; Ebbesen,~T.~W. Modifying Woodward--Hoffmann
  stereoselectivity under vibrational strong coupling. \emph{Angew. Chem., Int.
  Ed.} \textbf{2021}, \emph{60}, 5712--5717\relax
\mciteBstWouldAddEndPuncttrue
\mciteSetBstMidEndSepPunct{\mcitedefaultmidpunct}
{\mcitedefaultendpunct}{\mcitedefaultseppunct}\relax
\EndOfBibitem
\bibitem[Herrera and Spano(2016)Herrera, and Spano]{herrera2016cavity}
Herrera,~F.; Spano,~F.~C. Cavity-controlled chemistry in molecular ensembles.
  \emph{Phys. Rev. Lett.} \textbf{2016}, \emph{116}, 238301\relax
\mciteBstWouldAddEndPuncttrue
\mciteSetBstMidEndSepPunct{\mcitedefaultmidpunct}
{\mcitedefaultendpunct}{\mcitedefaultseppunct}\relax
\EndOfBibitem
\bibitem[Lather \latin{et~al.}(2019)Lather, Bhatt, Thomas, Ebbesen, and
  George]{lather2019cavity}
Lather,~J.; Bhatt,~P.; Thomas,~A.; Ebbesen,~T.~W.; George,~J. Cavity catalysis
  by cooperative vibrational strong coupling of reactant and solvent molecules.
  \emph{Angew. Chem., Int. Ed.} \textbf{2019}, \emph{58}, 10635--10638\relax
\mciteBstWouldAddEndPuncttrue
\mciteSetBstMidEndSepPunct{\mcitedefaultmidpunct}
{\mcitedefaultendpunct}{\mcitedefaultseppunct}\relax
\EndOfBibitem
\bibitem[Sidler \latin{et~al.}(2020)Sidler, Sch{\"a}fer, Ruggenthaler, and
  Rubio]{sidler2020polaritonic}
Sidler,~D.; Sch{\"a}fer,~C.; Ruggenthaler,~M.; Rubio,~A. Polaritonic chemistry:
  Collective strong coupling implies strong local modification of chemical
  properties. \emph{J. Phys. Chem. Lett.} \textbf{2020}, \emph{12},
  508--516\relax
\mciteBstWouldAddEndPuncttrue
\mciteSetBstMidEndSepPunct{\mcitedefaultmidpunct}
{\mcitedefaultendpunct}{\mcitedefaultseppunct}\relax
\EndOfBibitem
\bibitem[Li \latin{et~al.}(2021)Li, Nitzan, and Subotnik]{li2021collective}
Li,~T.~E.; Nitzan,~A.; Subotnik,~J.~E. Collective Vibrational Strong Coupling
  Effects on Molecular Vibrational Relaxation and Energy Transfer: Numerical
  Insights via Cavity Molecular Dynamics Simulations. \emph{Angew. Chem., Int.
  Ed.} \textbf{2021}, \emph{60}, 15533--15540\relax
\mciteBstWouldAddEndPuncttrue
\mciteSetBstMidEndSepPunct{\mcitedefaultmidpunct}
{\mcitedefaultendpunct}{\mcitedefaultseppunct}\relax
\EndOfBibitem
\bibitem[G{\'o}mez and Vendrell(2023)G{\'o}mez, and
  Vendrell]{gomez2023vibrational}
G{\'o}mez,~J.~A.; Vendrell,~O. Vibrational Energy Redistribution and
  Polaritonic Fermi Resonances in the Strong Coupling Regime. \emph{J. Phys.
  Chem. A} \textbf{2023}, \emph{127}, 1598--1608\relax
\mciteBstWouldAddEndPuncttrue
\mciteSetBstMidEndSepPunct{\mcitedefaultmidpunct}
{\mcitedefaultendpunct}{\mcitedefaultseppunct}\relax
\EndOfBibitem
\bibitem[Du \latin{et~al.}(2023)Du, Poh, and Yuen-Zhou]{du2023vibropolaritonic}
Du,~M.; Poh,~Y.~R.; Yuen-Zhou,~J. Vibropolaritonic Reaction Rates in the
  Collective Strong Coupling Regime: Pollak--Grabert--H{\"a}nggi Theory.
  \emph{J. Phys. Chem. C} \textbf{2023}, \emph{127}, 5230--5237\relax
\mciteBstWouldAddEndPuncttrue
\mciteSetBstMidEndSepPunct{\mcitedefaultmidpunct}
{\mcitedefaultendpunct}{\mcitedefaultseppunct}\relax
\EndOfBibitem
\bibitem[Simpkins \latin{et~al.}(2021)Simpkins, Dunkelberger, and
  Owrutsky]{simpkins2021mode}
Simpkins,~B.~S.; Dunkelberger,~A.~D.; Owrutsky,~J.~C. Mode-specific chemistry
  through vibrational strong coupling (or A wish come true). \emph{J. Phys.
  Chem. C} \textbf{2021}, \emph{125}, 19081--19087\relax
\mciteBstWouldAddEndPuncttrue
\mciteSetBstMidEndSepPunct{\mcitedefaultmidpunct}
{\mcitedefaultendpunct}{\mcitedefaultseppunct}\relax
\EndOfBibitem
\bibitem[Semenov and Nitzan(2019)Semenov, and Nitzan]{semenov2019electron}
Semenov,~A.; Nitzan,~A. Electron transfer in confined electromagnetic fields.
  \emph{J. Chem. Phys.} \textbf{2019}, \emph{150}, 174122\relax
\mciteBstWouldAddEndPuncttrue
\mciteSetBstMidEndSepPunct{\mcitedefaultmidpunct}
{\mcitedefaultendpunct}{\mcitedefaultseppunct}\relax
\EndOfBibitem
\bibitem[Wei and Hsu(2022)Wei, and Hsu]{wei2022cavity}
Wei,~Y.-C.; Hsu,~L.-Y. Cavity-Free Quantum-Electrodynamic Electron Transfer
  Reactions. \emph{J. Phys. Chem. Lett.} \textbf{2022}, \emph{13},
  9695--9702\relax
\mciteBstWouldAddEndPuncttrue
\mciteSetBstMidEndSepPunct{\mcitedefaultmidpunct}
{\mcitedefaultendpunct}{\mcitedefaultseppunct}\relax
\EndOfBibitem
\bibitem[Gruner and Welsch(1996)Gruner, and Welsch]{gruner1996green}
Gruner,~T.; Welsch,~D.-G. Green-function approach to the radiation-field
  quantization for homogeneous and inhomogeneous Kramers-Kronig dielectrics.
  \emph{Phys. Rev. A} \textbf{1996}, \emph{53}, 1818\relax
\mciteBstWouldAddEndPuncttrue
\mciteSetBstMidEndSepPunct{\mcitedefaultmidpunct}
{\mcitedefaultendpunct}{\mcitedefaultseppunct}\relax
\EndOfBibitem
\bibitem[Dung \latin{et~al.}(1998)Dung, Kn{\"o}ll, and Welsch]{dung1998three}
Dung,~H.~T.; Kn{\"o}ll,~L.; Welsch,~D.-G. Three-dimensional quantization of the
  electromagnetic field in dispersive and absorbing inhomogeneous dielectrics.
  \emph{Phys. Rev. A} \textbf{1998}, \emph{57}, 3931\relax
\mciteBstWouldAddEndPuncttrue
\mciteSetBstMidEndSepPunct{\mcitedefaultmidpunct}
{\mcitedefaultendpunct}{\mcitedefaultseppunct}\relax
\EndOfBibitem
\bibitem[Buhmann(2013)]{buhmann2013dispersion}
Buhmann,~S.~Y. \emph{Dispersion Forces I: Macroscopic quantum electrodynamics
  and ground-state Casimir, Casimir--Polder and van der Waals forces};
  Springer, 2013\relax
\mciteBstWouldAddEndPuncttrue
\mciteSetBstMidEndSepPunct{\mcitedefaultmidpunct}
{\mcitedefaultendpunct}{\mcitedefaultseppunct}\relax
\EndOfBibitem
\bibitem[Wei and Hsu(2023)Wei, and Hsu]{wei2023polaritonic}
Wei,~Y.-C.; Hsu,~L.-Y. Polaritonic Huang--Rhys Factor: Basic Concepts and
  Quantifying Light--Matter Interactions in Media. \emph{J. Phys. Chem. Lett.}
  \textbf{2023}, \emph{14}, 2395--2401\relax
\mciteBstWouldAddEndPuncttrue
\mciteSetBstMidEndSepPunct{\mcitedefaultmidpunct}
{\mcitedefaultendpunct}{\mcitedefaultseppunct}\relax
\EndOfBibitem
\bibitem[Marcus(1993)]{marcus1993electron}
Marcus,~R.~A. Electron transfer reactions in chemistry. Theory and experiment.
  \emph{Rev. Mod. Phys.} \textbf{1993}, \emph{65}, 599\relax
\mciteBstWouldAddEndPuncttrue
\mciteSetBstMidEndSepPunct{\mcitedefaultmidpunct}
{\mcitedefaultendpunct}{\mcitedefaultseppunct}\relax
\EndOfBibitem
\bibitem[Wang \latin{et~al.}(2020)Wang, Scholes, and Hsu]{wang2020coherent}
Wang,~S.; Scholes,~G.~D.; Hsu,~L.-Y. Coherent-to-incoherent transition of
  molecular fluorescence controlled by surface plasmon polaritons. \emph{J.
  Phys. Chem. Lett.} \textbf{2020}, \emph{11}, 5948--5955\relax
\mciteBstWouldAddEndPuncttrue
\mciteSetBstMidEndSepPunct{\mcitedefaultmidpunct}
{\mcitedefaultendpunct}{\mcitedefaultseppunct}\relax
\EndOfBibitem
\bibitem[Lee and Hsu(2020)Lee, and Hsu]{lee2020controllable}
Lee,~M.-W.; Hsu,~L.-Y. Controllable frequency dependence of resonance energy
  transfer coupled with localized surface plasmon polaritons. \emph{J. Phys.
  Chem. Lett.} \textbf{2020}, \emph{11}, 6796--6804\relax
\mciteBstWouldAddEndPuncttrue
\mciteSetBstMidEndSepPunct{\mcitedefaultmidpunct}
{\mcitedefaultendpunct}{\mcitedefaultseppunct}\relax
\EndOfBibitem
\bibitem[Wei \latin{et~al.}(2021)Wei, Lee, Chou, Scholes, Schatz, and
  Hsu]{wei2021can}
Wei,~Y.-C.; Lee,~M.-W.; Chou,~P.-T.; Scholes,~G.~D.; Schatz,~G.~C.; Hsu,~L.-Y.
  Can Nanocavities Significantly Enhance Resonance Energy Transfer in a Single
  Donor--Acceptor Pair? \emph{J. Phys. Chem. C} \textbf{2021}, \emph{125},
  18119--18128\relax
\mciteBstWouldAddEndPuncttrue
\mciteSetBstMidEndSepPunct{\mcitedefaultmidpunct}
{\mcitedefaultendpunct}{\mcitedefaultseppunct}\relax
\EndOfBibitem
\bibitem[Buck \latin{et~al.}(2019)Buck, Wilson, and
  Mani]{buck2019intramolecular}
Buck,~J.~T.; Wilson,~R.~W.; Mani,~T. Intramolecular long-range charge-transfer
  emission in donor--bridge--acceptor systems. \emph{J. Phys. Chem. Lett.}
  \textbf{2019}, \emph{10}, 3080--3086\relax
\mciteBstWouldAddEndPuncttrue
\mciteSetBstMidEndSepPunct{\mcitedefaultmidpunct}
{\mcitedefaultendpunct}{\mcitedefaultseppunct}\relax
\EndOfBibitem
\bibitem[Vasa and Lienau(2018)Vasa, and Lienau]{vasa2018strong}
Vasa,~P.; Lienau,~C. Strong light--matter interaction in quantum emitter/metal
  hybrid nanostructures. \emph{ACS Photonics} \textbf{2018}, \emph{5},
  2--23\relax
\mciteBstWouldAddEndPuncttrue
\mciteSetBstMidEndSepPunct{\mcitedefaultmidpunct}
{\mcitedefaultendpunct}{\mcitedefaultseppunct}\relax
\EndOfBibitem
\bibitem[Rivera and Kaminer(2020)Rivera, and Kaminer]{rivera2020light}
Rivera,~N.; Kaminer,~I. Light--matter interactions with photonic
  quasiparticles. \emph{Nat. Rev. Phys} \textbf{2020}, \emph{2}, 538--561\relax
\mciteBstWouldAddEndPuncttrue
\mciteSetBstMidEndSepPunct{\mcitedefaultmidpunct}
{\mcitedefaultendpunct}{\mcitedefaultseppunct}\relax
\EndOfBibitem
\bibitem[Purcell(1946)]{purcell1946proceedings}
Purcell,~E.~M. Proceedings of the American Physical Society, b10. Spontaneous
  emission probabilities at radio frequencies. \emph{Phys. Rev.} \textbf{1946},
  \emph{69}, 674\relax
\mciteBstWouldAddEndPuncttrue
\mciteSetBstMidEndSepPunct{\mcitedefaultmidpunct}
{\mcitedefaultendpunct}{\mcitedefaultseppunct}\relax
\EndOfBibitem
\bibitem[Vahala(2003)]{vahala2003optical}
Vahala,~K.~J. Optical microcavities. \emph{Nature} \textbf{2003}, \emph{424},
  839--846\relax
\mciteBstWouldAddEndPuncttrue
\mciteSetBstMidEndSepPunct{\mcitedefaultmidpunct}
{\mcitedefaultendpunct}{\mcitedefaultseppunct}\relax
\EndOfBibitem
\end{mcitethebibliography}

\end{document}